\documentclass[preprintnumbers,prb,amsmath,amssymb]{revtex4}
\usepackage{graphicx}
\usepackage{bm}

\begin{document}

\title{
Proton energy behavior by variation of the target density in laser acceleration
}

\author{Toshimasa~Morita}

\affiliation{Quantum Beam Science Research Directorate,
National Institutes for Quantum and Radiological Science and Technology,
8-1-7 Umemidai, Kizugawa, Kyoto 619-0215, Japan}

\date{\today}

\begin{abstract}
Ion acceleration using a laser pulse irradiating a thin disk target is examined
using three-dimensional and two-dimensional particle-in-cell simulations.
A laser pulse of $620$ TW, with an intensity of $5\times 10^{21}$ W/cm$^{2}$
and a duration time of $27$ fs is irradiated on a double-layer target.
Simulations are performed by varying the ion density, i.e., electron density,
of the first layer.
It is shown that the obtained proton energy jumps at a certain density of the
first layer, which is made of a ``light'' material such as carbon;
that is,
Coulomb explosion of the target and radiation pressure acceleration
act effectively above a certain density.
Moreover, even at the same electron density, the reflection of the laser pulse
from the first layer is small for a ``light'' material.
\end{abstract}

\maketitle

\section{INTRODUCTION}
The recent progress of compact laser systems has been remarkable;
their power and intensity have been significantly improved,
and a high contrast ratio has been achieved.
Laser ion acceleration is one attractive application of this technology
\cite{DNP,BWE}.
If sufficiently high energy ions are generated by a compact laser system,
compact accelerators at a low cost may become feasible.
Realization of such systems may contribute significantly to scientific
research and technology development in the future.
However, the ion energies that have currently been achieved by laser ion
acceleration are insufficient for some applications such as hadron therapy
\cite{CLK,SNV}.
In addition,
such applications require the ion beam to have a narrow energy spread
with a sufficient number of ions \cite{ROT,ESI,BEE}.
One simple way to obtain high energy ions by laser acceleration is
to use a high power laser.
However, such laser systems are both expensive and not compact.
Therefore, it is important to study how to generate ions with
the highest energy possible using a certain laser capacity
\cite{BWP,DL,HSM,PPM,PRK,SPJ,Toncian,TM1,TM2}.

In this paper,
I investigate a way to generate higher energy ions,
moreover a high quality, small energy spread, ion beam.
A double-layer target is used since
it can generate a high quality proton beam \cite{DL}.
The double-layer target is composed of a thin and low density hydrogen layer
on a high-$Z$ atom layer (see Fig. \ref{fig:fig-1}(a)).
This type of target could be made by some processing technologies.
Although such a target may be difficult to prepare, a doped target will
become one automatically in the early acceleration stage \cite{TM2}.
The higher energy protons can be obtained using the ``light'' material for
the target \cite{TM1}.
Here, ``light'' means small $m/q$, where $m$ is mass and
$q$ is the charge of an ion, and ``heavy'' denotes large $m/q$.
In this paper,
I examine a way to produce higher energy ions by
coupling strong Coulomb explosion of the target,
powerful radiation pressure acceleration (RPA),
and efficient laser energy absorption as acceleration mechanisms.

The total potential energy of the ions in the target grows
with the increase of the target density, since higher density mean
the distance between the ions in the target are shorter.
Therefore,
strong Coulomb explosion occurs in high density targets.
In addition, in the RPA, the high density of ions and electrons is considered
a good effect for this scheme, because it is important to sharply spatially
separate high density ions and electrons.
With these considerations, we should be able to obtain high energy protons
using the high density target.
On the other hand,
the electron number density in the target also increases along with
the ion number density.
Then, the negative effect of the reflection of a laser pulse
on a target surface increases along with target density.
Therefore,
there is an appropriate density for generating high energy ions
that can be determined by managing the tradeoff between the positive effects
of strong Coulomb explosion and PRA
and the negative effect of reflected laser energy.

To examine the dependence of the achieved proton energy on the density of
the target, we performed three- and two-dimensional (3D and 2D, respectively)
particle-in-cell (PIC) simulations.
As the ``light'' material for the first layer, carbon is employed.
Gold is also used as a ``heavy'' material for comparison.

In Sec. \ref{para}, the simulation parameters are shown.
In Sec. \ref{sim}, I present the dependence of the proton energy on
the first layer density and show
that high energy protons can be generated by certain ion acceleration schemes,
namely strong Coulomb explosion, strong RPA,
and less laser reflection.
Analytical consideration is presented in Sec. \ref{analy}.

\section{SIMULATION PARAMETERS} \label{para}

The simulations were performed with a parallel electromagnetic code
based on the PIC method \cite{CBL}.
The parameters used in the simulations are shown in below.

I use an idealized model in which a Gaussian linear polarized laser pulse is
normally incident on a double-layer mass-limited target.
The laser pulse with dimensionless amplitude
$a_0=q_eE/m_{e}\omega c=50$, corresponding to the laser peak intensity of
$5\times 10^{21}$ W/cm$^{2}$, has a $27$ fs (FWHM) duration
and is focused on a spot of size $3.2 \mu$m (FWHM),
corresponding to a laser power of $620$ TW and a laser energy of $18$ J.
The laser wavelength is $\lambda = 0.8 \mu $m.

In the 3D simulation cases,
the number of grid cells is equal to $5620\times 3888\times 3888$
along the $X$, $Y$, and $Z$-axes.
Correspondingly, the simulation box size is
$80 \mu$m $\times 55 \mu$m $\times 55 \mu$m.
The laser propagation direction is set to be along the $X$ direction,
and the electric field is oriented in the $Y$ direction.
The boundary conditions for the particles and the fields are
periodic in the transverse ($Y$,$Z$) directions and absorbing at the
boundaries of the computation box along the $X$-axis.
The front side surface of the first layer is placed at $X=32 \mu$m,
and the center of the laser pulse is located $16 \mu$m behind it. 
The total number of quasiparticles is $1.3 \times 10^{8}$.
In the 2D simulation cases,
the number of grid cells is equal to $20000\times20000$ along the $X$ and
$Y$-axes, and the simulation box size is $127 \mu$m $\times 127 \mu$m.
The front side surface of the first layer location is $X=56 \mu$m,
and the center of the laser pulse is located $16 \mu$m behind it.
The total number of quasiparticles is $1.3 \times 10^{7}$.
These parameters described above are used for all simulations in this paper.

\begin{figure}[tbp]
\includegraphics[clip,width=7.0cm,bb=9 3 399 490]{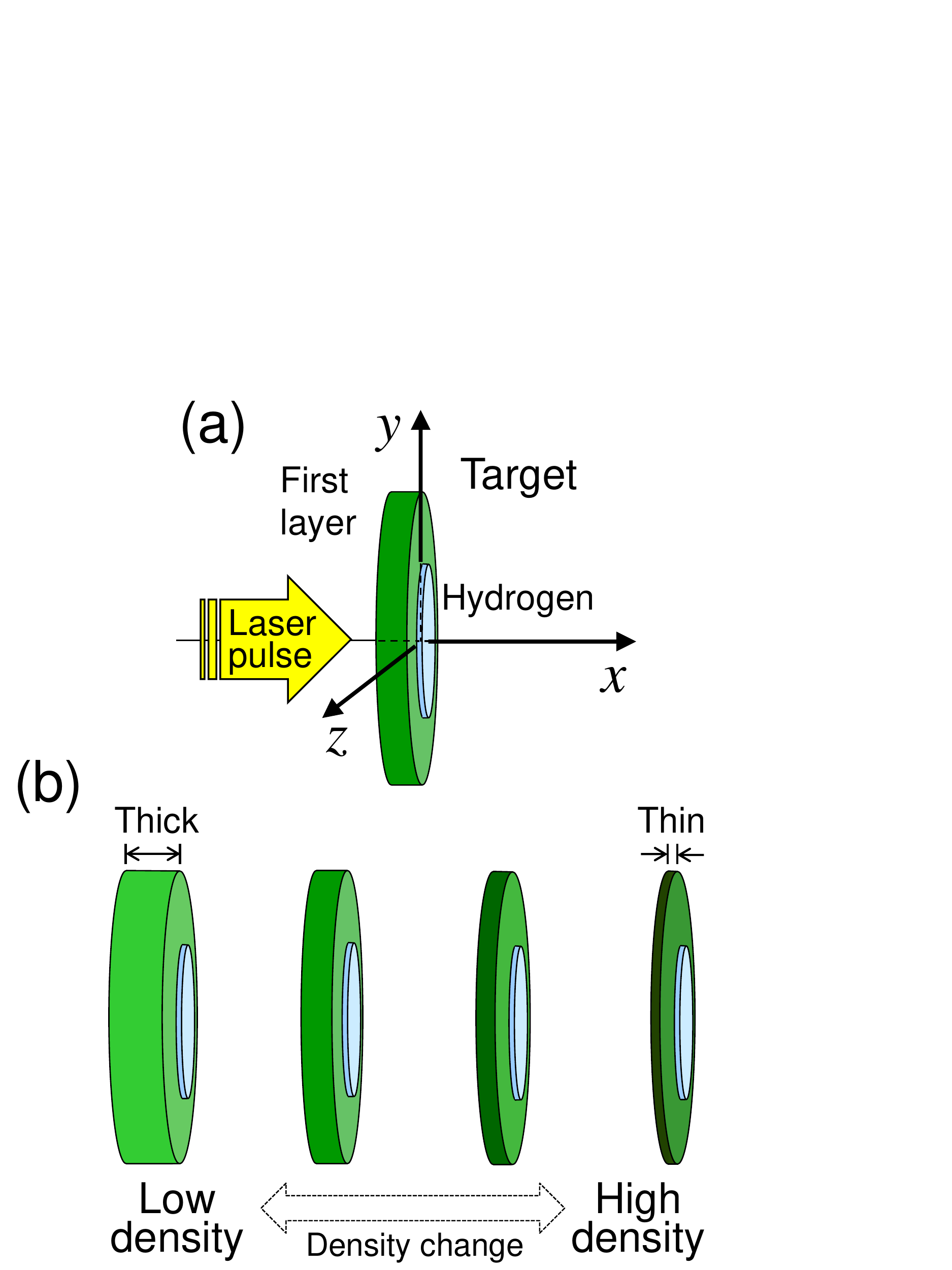}
\caption{
(a) Laser pulse is normally incident on a double-layer target.
(b) The ion, i.e., electron, density inside the first layer is varied.
The ion, electron, density of the first layer is changed by adjusting
the layer thickness, maintaining the same number of ions and electrons.
}
\label{fig:fig-1}
\end{figure}

In this paper,
the simulation results using different materials in the first layer are shown.
One consists of carbon (a ``light'' material) and
the other of gold (a ``heavy'' material).
The simulation results at different ion densities, i.e., electron densities,
in each material at the first layer of the target are shown.
The number of ions, i.e., electrons, is the same in all cases,
even under different densities.
The density is changed by adjusting the target thickness, i.e.,
we define the high density targets by narrowing the thickness of
the first layer while maintaining the same number of ions and electrons
(see Fig. \ref{fig:fig-1}(b)).
By defining it this way,
when the same number electrons are pulled out from the target,
the surface charge density is the same in all cases,
although the ion and electron densities differ.
When the surface charge density is the same,
the acceleration field is the same (see Eq. (\ref{exx})).
Therefore, we can evaluate only the target density effect.

Both layers of the double-layer target are disk shaped.
The first layer has a diameter of $6.4\mu$m, and its thickness is varied
in different density cases, having values of $0.4\mu$m in $30 n_\mathrm{cr}$,
$0.15\mu$m in $80 n_\mathrm{cr}$, $0.04\mu$m in $300 n_\mathrm{cr}$ and so on,
where $n_\mathrm{cr}=\epsilon_0m_e\omega^2/q_e^2$ is the critical density,
$n_\mathrm{cr}=1.7\times 10^{21}$ cm$^{-3}$.
The ionization state of each ion in the first layer is assumed to be $Z_{i}=+6$.
That is, the carbons are fully striped.
The gold is assumed to have the same ionization state as the carbon
for the purpose of the comparison case.
The second, hydrogen, layer is narrower and thinner;
its diameter is $3.2 \mu$m and thickness is $0.024 \mu$m.
The electron density
inside the hydrogen layer is $n_{e}=9\times 10^{20}$ cm$^{-3}$.

In the text and figures of this paper, we use $xyz-$coordinates.
The $x$-axis denotes the direction perpendicular to the target surface,
and the $y$ and $z$-axes are parallel to the target surface.
That is, the directions of the $x$, $y$, and $z$-axes are the same as those of
the $X$, $Y$, and $Z$-axes, respectively. 
The origin in $xyz-$coordinates is located at the center of
the rear surface of the initial first layer.

\section{PROTON ENERGY VS. ION DENSITY} \label{sim}

\begin{figure}[tbp]
\includegraphics[clip,width=11.0cm,bb=0 7 437 316]{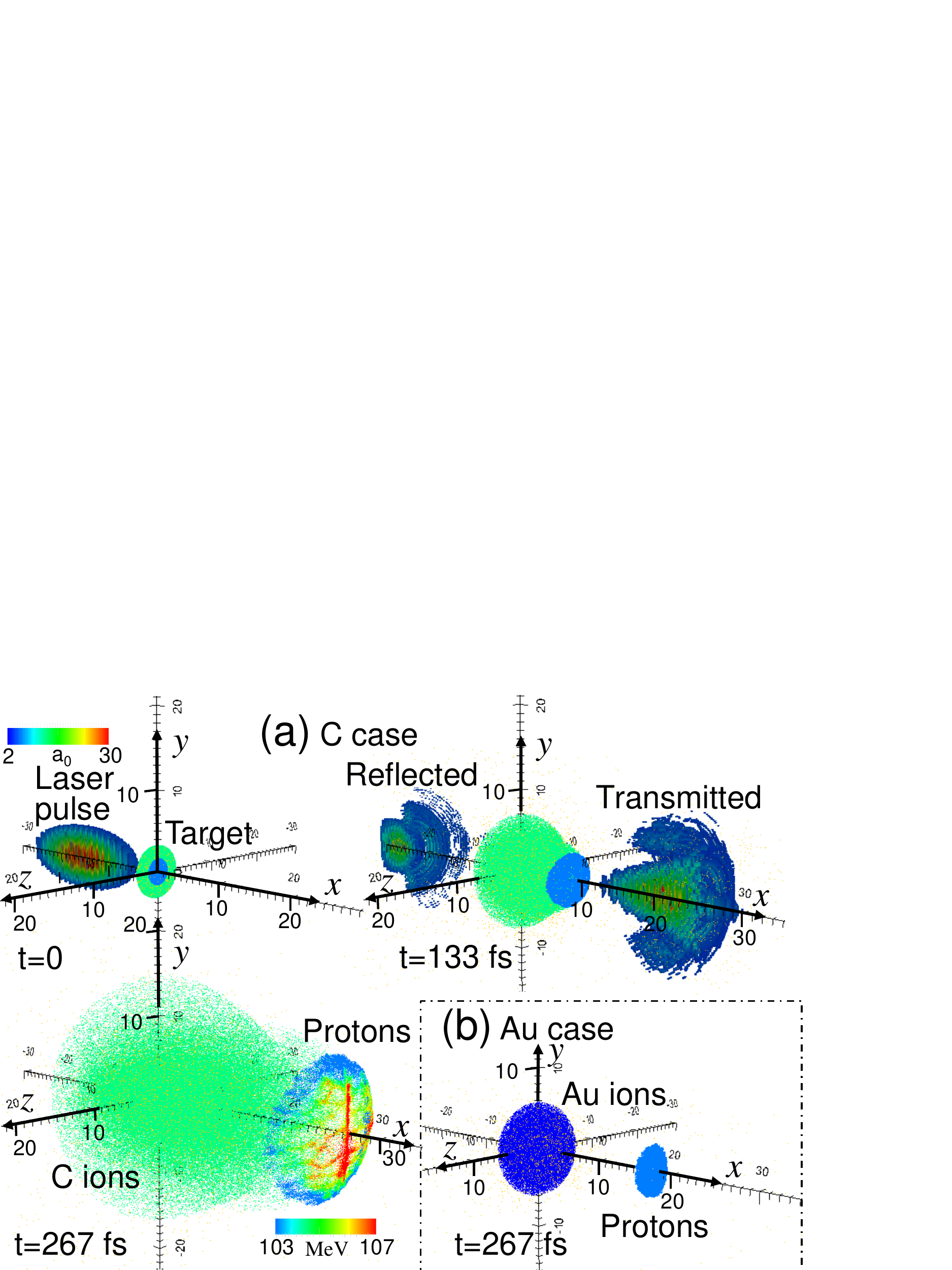}
\caption{
(a) Carbon case in which initial electron density is $60 n_\mathrm{cr}$.
A 3D view is presented of the particle distribution and electric field magnitude
(isosurface for value $a_0=2$) at times $t=0$, $133$, and $267$ fs.
Half of the electric field box has been excluded to reveal the internal
structure. For protons, the color corresponds to their energy.
(b) Gold case at $t=267$ fs, for which the initial electron density is
the same as in the carbon case.
}
\label{fig:fig-2}
\end{figure}

In this section, the dependence of the proton energy on the
ion density of the first layer is described.
Figure \ref{fig:fig-2} shows the particle distribution and electric field
magnitude of the $60 n_\mathrm{cr}$ carbon and gold cases in a 3D simulation.
In the carbon case, the states at times $t=0$, $133$ and $267$ fs are shown.
After strong interactions with the target,
part of the laser pulse is reflected while another part is transmitted
(see Fig. \ref{fig:fig-2}(a) $t=133$ fs).
We see that the carbon ions are distributed over a much wider area
by the Coulomb explosion.
The obtained proton energy of the carbon case is much higher than
that of the gold case.
The maximum proton energy, $\mathcal{E}$, is $110$ MeV for the carbon case
and $49$ MeV for the gold case at $t=267$ fs.
The gold ions are distributed over a much smaller area than the carbon ions.
This means the Coulomb explosion of the gold target is much weaker than
that of the carbon target even for the same electrons and ions density.

\begin{figure}[tbp]
\includegraphics[clip,width=7.5cm,bb=3 0 431 404]{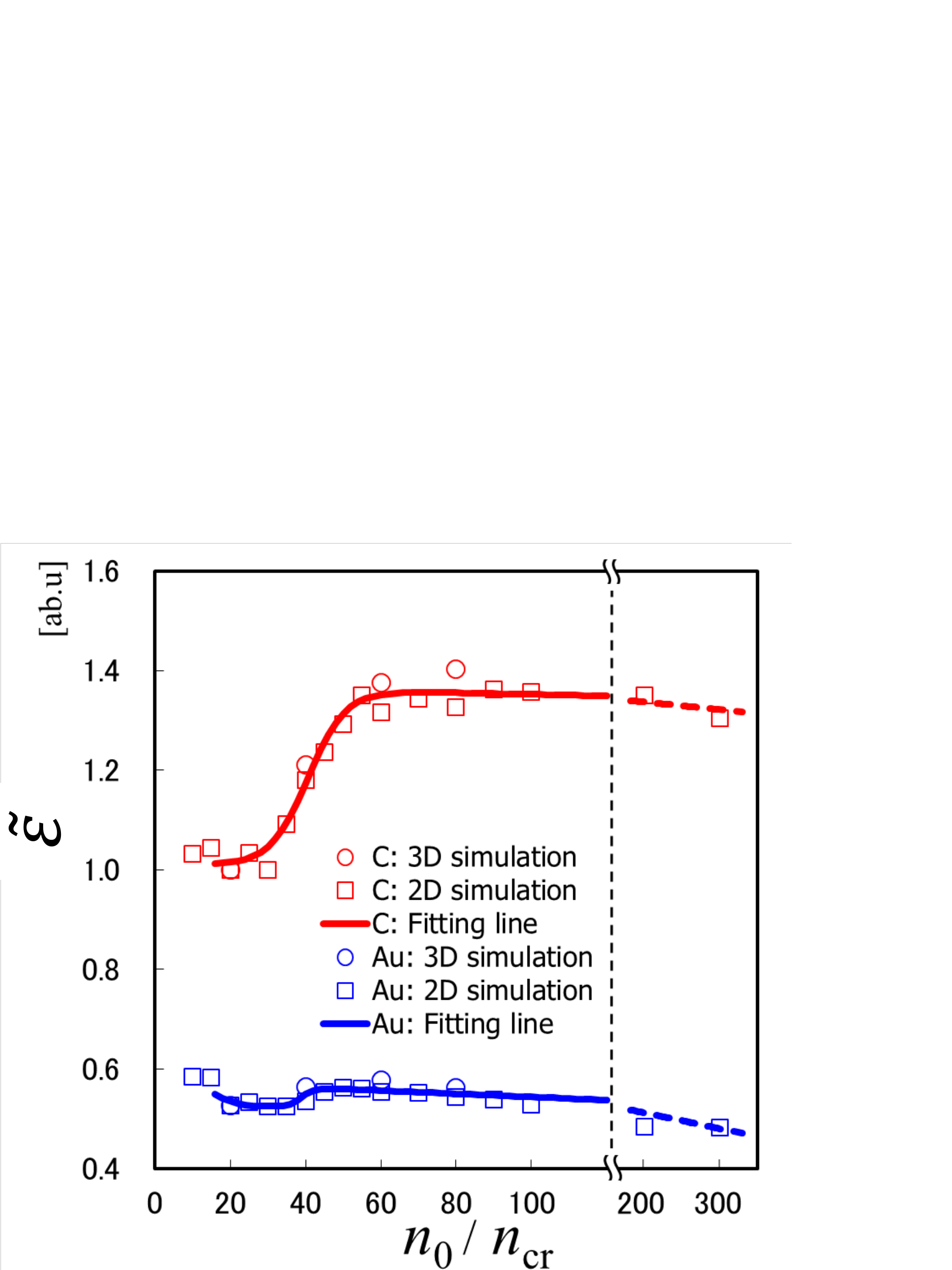}
\caption{
Proton energy, $\tilde{\mathcal{E}}$,
normalized by its minimum in the carbon case,
versus the initial electron density, $n_0$, of the first layer
at $t=267$ fs.
The generated proton energy jumps as density changes from
30$ n_\mathrm{cr}$ to 60$ n_\mathrm{cr}$ in the carbon case
and is much higher than in the gold case.
}
\label{fig:fig-3}
\end{figure}

\begin{figure}[tbp]
\includegraphics[clip,width=11.0cm,bb=3 0 520 240]{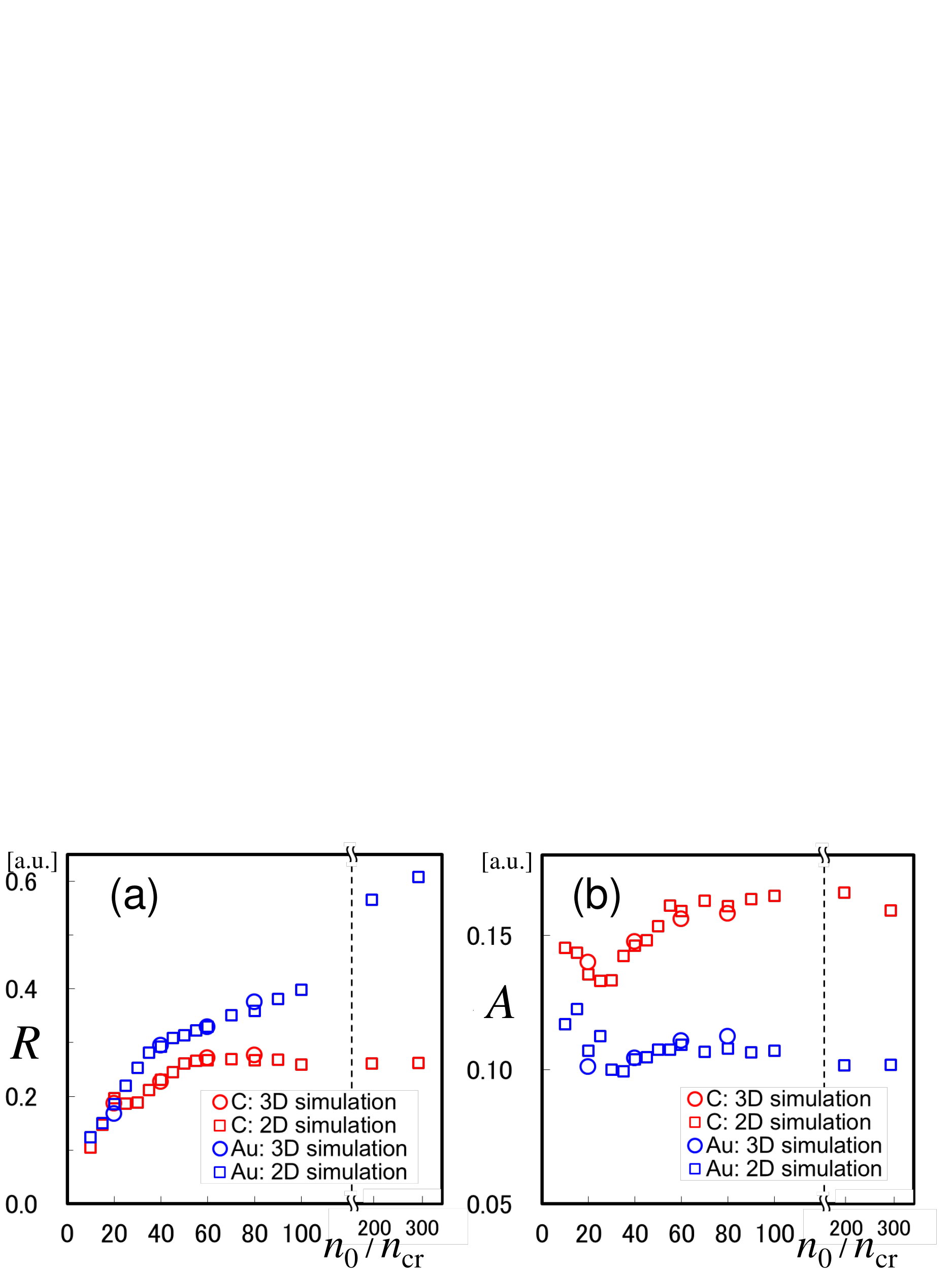}
\caption{
(a) Reflected laser energy, $R$, from the target surface and
(b) absorbed laser energy, $A$, in the target
versus the initial electron density of the first layer.
These energies are presented as ratios of the initial laser pulse energy.
}
\label{fig:fig-5}
\end{figure}

\begin{figure}[tbp]
\includegraphics[clip,width=10.0cm,bb=8 0 486 279]{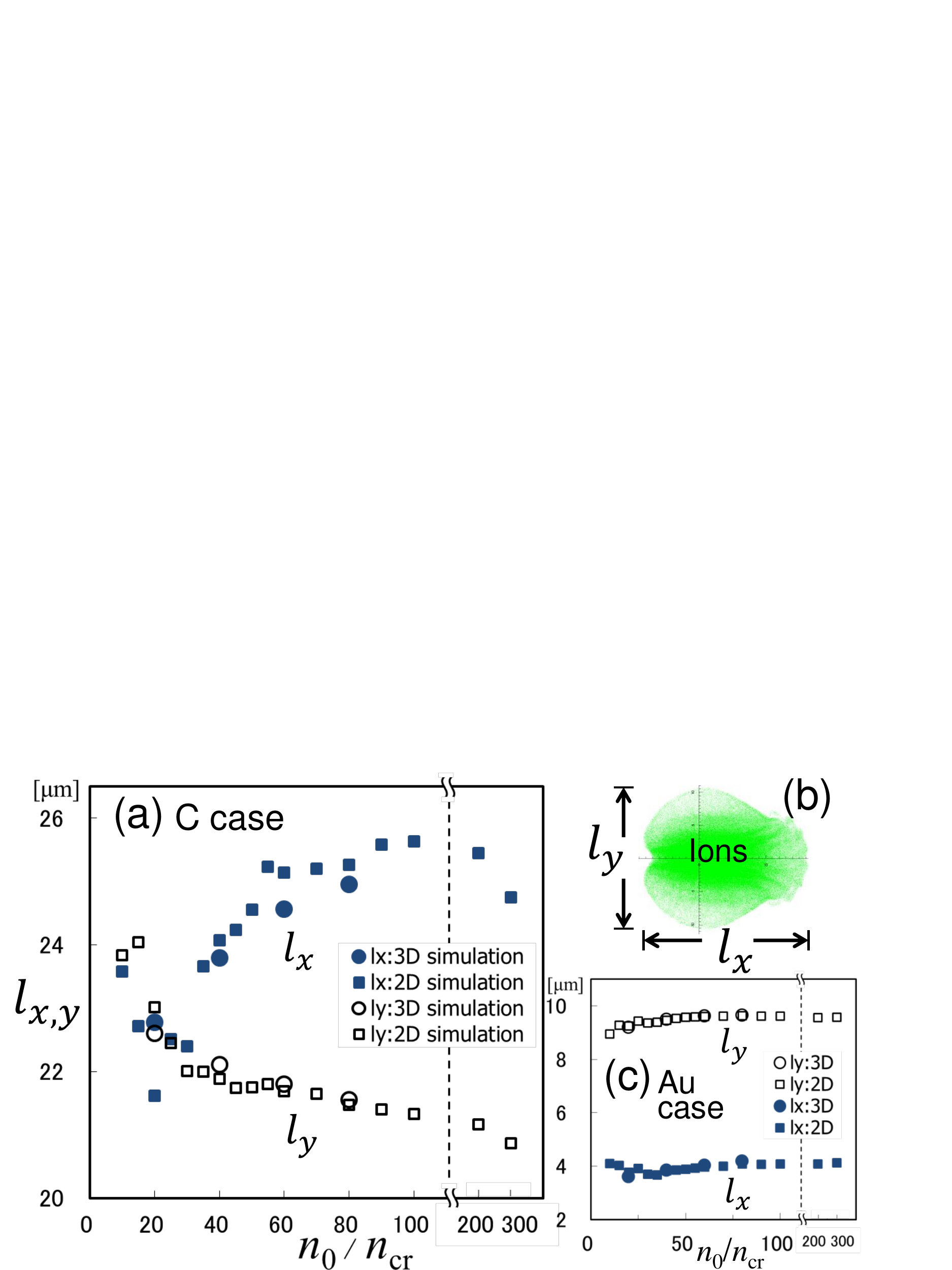}
\caption{
The $x$ and $y$ direction lengths, $\ell_x$ and $\ell_y$, of the ion cloud
of the
first layer in the carbon (a) and gold (c) cases at $t=200$ fs as a function
of first layer density. The $\ell_x$ of the carbon case increases along
with first layer density.
(b) $\ell_x$ and $\ell_y$ are defined by the maximum $-$ minimum
coordinates of ions.
}
\label{fig:fig-11}
\end{figure}

The obtained proton energy, $\tilde{\mathcal{E}}$,
which is normalized by its minimum in the carbon case,
is shown as a function of the electron density of the target,
$n_0$, in Fig. \ref{fig:fig-3}.
Where $n_0$ is the initial electron density in the first layer.
In the carbon case, we see that the proton energy jumps up in
from $30 n_\mathrm{cr}$ to $60 n_\mathrm{cr}$
and that it is almost flat in above $60 n_\mathrm{cr}$.
On the other hand, in the gold case,
there is no such dramatic changing of the proton energy.
Rather, it decreases almost monotonously.
The proton energy is normalized by the minimal value for the carbon case
in each dimensional case,
i.e.,
$\tilde{\mathcal{E}}=\mathcal{E}/\mathcal{E}_\mathrm{C,20}$,
where $\mathcal{E}_\mathrm{C,20}$ is the proton energy in the $20 n_\mathrm{cr}$
carbon case, which is the minimum value.
In addition, only in the 2D gold case, those are adjusted to the 3D values
by a multiplicative factor of
$\tilde{\mathcal{E}}_\mathrm{3D,Au,20}/\tilde{\mathcal{E}}_\mathrm{2D,Au,20}$,
where $\tilde{\mathcal{E}}_\mathrm{3D,Au,20}$ is the value of the 3D gold case 
at $20n_\mathrm{cr}$ and
$\tilde{\mathcal{E}}_\mathrm{2D,Au,20}$ is the value of the 2D one.
In general,
the proton energy obtained by the 2D PIC simulation is much higher than
that obtained by
the 3D PIC simulation depending on the characteristics of the 2D model.
Of course, 3D simulation results are much more accurate.
Therefore, 
the 2D simulation result must be adjusted to the 3D value,
especially when it is discussed with 3D simulation results.
A similar adjustment for the 2D simulation results is performed
in Figs. \ref{fig:fig-5} and \ref{fig:fig-11}.

Below, I show why the proton energy jump appears in the carbon,
``light'' material, target, and
why the obtained proton energies in all carbon cases
are higher than in any gold, ``heavy'' material, cases.

\begin{figure}[tbp]
\includegraphics[clip,width=7.0cm,bb=6 9 268 600]{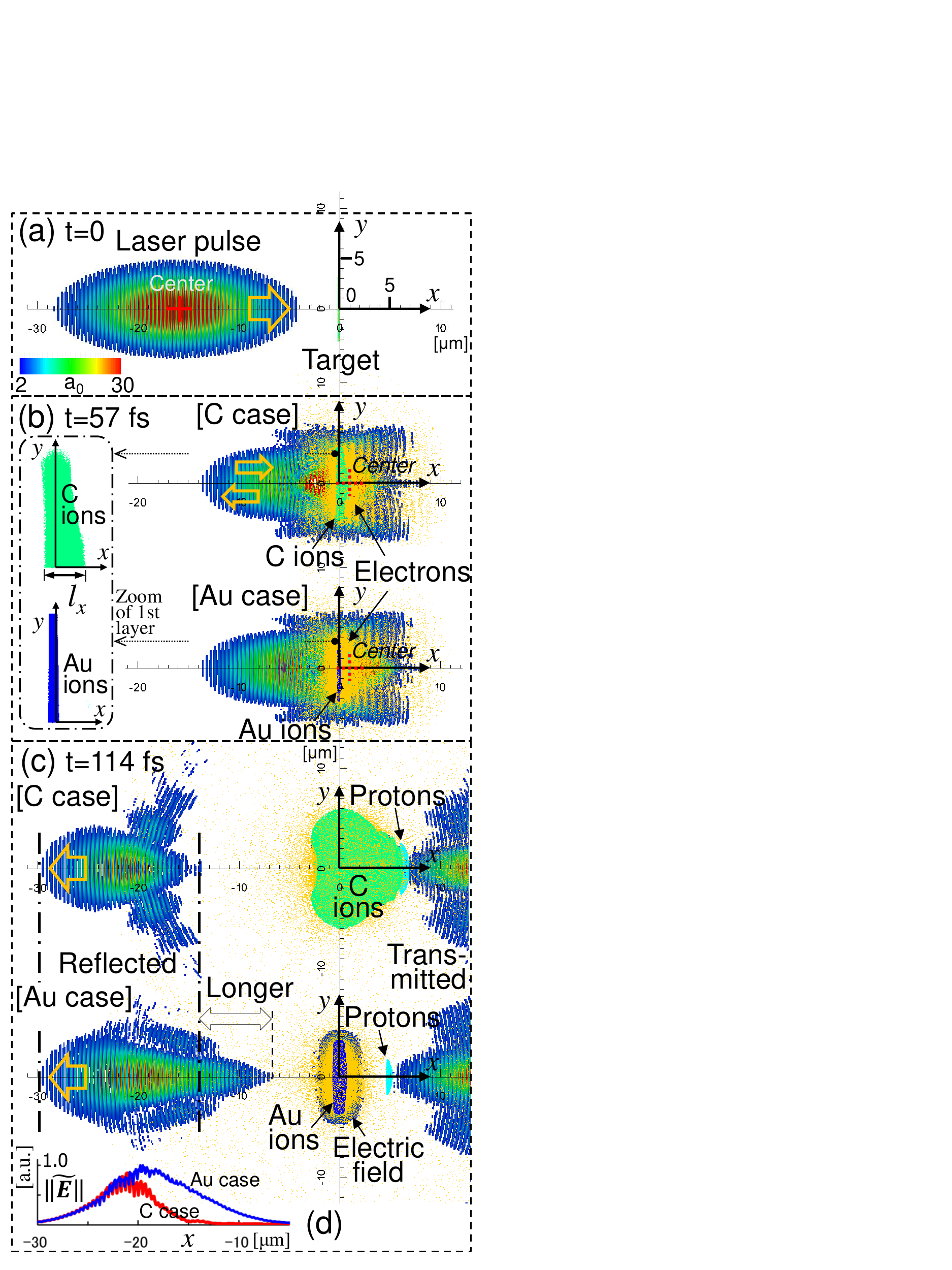}
\caption{
Spatial distribution of particles and electric field magnitude
(isosurface for value $a_0=2$)
of the $60 n_\mathrm{cr}$ carbon and gold targets in a 3D simulation.
A 2D projection of the section at $z=0$ is shown as viewed along the $z$-axis.
(a) The initial laser pulse and the target.
(b) The states in which of almost half of the laser pulse has interacted
with the target (t=$57$ fs).
The dot crosses show the points that would be the centers of the laser pulses
if they did not to interact with the target.
The reflected laser and the back half of the laser pulse are overlapped
in the electric field for $x<0$.
In the inset: zoom view of the first layer of the target.
The thickness, $\ell_x$, of the carbon target is much thicker than
that of the gold target.
(c) The states in time after interaction
of the laser and the target (t=$114$ fs).
The reflected laser pulse length of the gold case is longer than in
the carbon case.
(d) The electric field magnitude of the reflected laser in the carbon and
the gold cases at $t=114$ fs, which is normalized by its maximum value.
}
\label{fig:fig-4}
\end{figure}

Figure \ref{fig:fig-4} shows the electric field magnitude and the particle,
ions and electrons, distribution in the $60 n_\mathrm{cr}$
carbon and gold cases at early simulation times in the 3D simulations.
The $60 n_\mathrm{cr}$ result is chosen
since the jump in the obtained proton energy ends around this density.
The laser pulse and the target states are shown at t$=0$ (initial state),
$t=57$ fs (when almost half of the laser pulse has interacted with the target),
and
$t=114$ fs (when the laser pulse and the target interaction has ended).
At $t=57$ fs,
although almost half of the laser pulse has yet to interact with the target,
the thickness of the carbon target increases beyond that of the gold target
(see the inset in Fig. \ref{fig:fig-4}(b)).
This means the electron density, $n_\mathrm{e}$,
of the carbon target becomes much lower than the gold target,
since electrons are distributed nearly uniformly in the target,
because electrons move very quickly to follow the ions.
Thus,  laser reflection in the carbon case is much less than that in
the gold case at a later time.
We see this effect in the shape of reflected laser (Fig. \ref{fig:fig-4}(c)).
The tail of reflected laser of the carbon case
is shorter than that in the gold case.
However, the head of the reflected laser has almost the same shape
for both cases.
Figure \ref{fig:fig-4}(d) shows the distribution of the electric field
magnitude of the reflected laser on the center line (on $x$ axis).
The electric field magnitude, $\tilde{\| \bm{E} \|}$, is averaged
over the laser wavelength and then normalized by its maximum value
which is the value at $x=-20 \mu$m in the gold case.
The electric field magnitude in the head part, $-30 < x < -22 \mu$m,
is the same for the carbon and gold cases.
However, in the $-22 \mu$m $< x$ region,
the electric field in the carbon case decreases much more rapidly
than in the gold case and quickly becomes $\approx 0$.
The reflected laser energy from the carbon, ``light'' material, target
is thus found to be smaller than that from the gold, ``heavy'' material, target
even with the same density of electrons and ions.
At $t=114$ fs (see Fig. \ref{fig:fig-4}(c)),
the carbon target explodes strongly, especially in the $x$ direction, and
the center point of the carbon ion cloud moves in the $+x$ direction.
However,
the gold target does not experience such a strong explosion or movement.

The reflected, $R$, and absorbed, $A$, laser energies normalized
by the initial laser energy are shown as functions of the initial electron
density of the first layer in Fig. \ref{fig:fig-5}.
Although the reflected energies in the carbon and gold cases are almost the
same for $n_0 < 20 n_\mathrm{cr}$, the reflected energy from the carbon target
becomes smaller than that for the gold target in $n_0 \geq 20 n_\mathrm{cr}$.
In the gold target,
the reflected energy keeps rising with increasing target density.
However, in the carbon targets,
although it is also rising in $n_0 < 60 n_\mathrm{cr}$,
the reflected energies are almost the same for $ n_0 \geq 60 n_\mathrm{cr}$
(Fig. \ref{fig:fig-5}(a)).
For the ``light'' material target,
the laser reflection energy does not increase with
the target density above a certain value.
The absorbed energy of the carbon target is always higher than
that of the gold target (Fig. \ref{fig:fig-5}(b)).
In the gold case,
the absorbed energy roughly decreases with increasing target density.
This trend is similar to that in Fig. \ref{fig:fig-3}.
From only this gold case result,
we might think that low $n_0$ targets, e.g., gas targets,
are good for obtaining high energy ions.
However, in the carbon target,
the absorbed energy starts increasing around $30 n_\mathrm{cr}$
and keeps increasing to around $60 n_\mathrm{cr}$.
Then, they become almost the same values at in $n_0 \geq 60 n_\mathrm{cr}$.
It is shown that
the high density, $n_0 \geq 60 n_\mathrm{cr}$, target of ``light'' material
is better than the low density targets, gas targets,
for obtaining high energy ions.
The adjustment for 2D simulation results is also done
for the reason described above.
The 2D values are adjusted such that the average value corresponds to the
3D value in each material target.
The density values at $20, 40, 60,$ and $80 n_\mathrm{cr}$ are taken
as averages.

\begin{figure}[tbp]
\includegraphics[clip,width=11.0cm,bb=5 2 536 250]{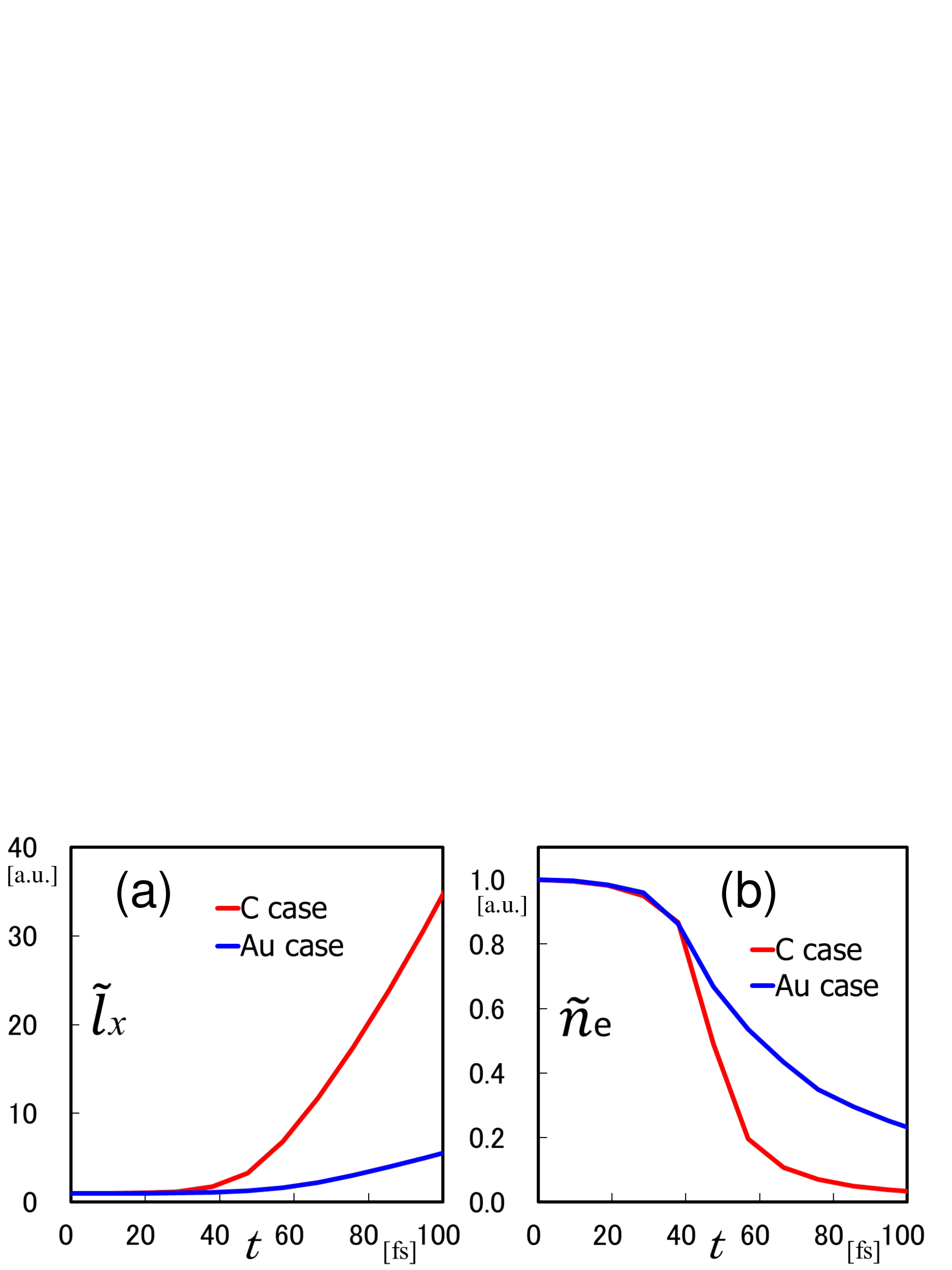}
\caption{
(a) Target thickness, $\tilde{\ell}_x$,
and (b) electron density, $\tilde{n}_\mathrm{e}$,
of the $60 n_\mathrm{cr}$ carbon and gold cases as functions of time.
These are normalized by their initial values.
The carbon target thickness becomes much ticker than that of the gold target
for $t>40$ fs and the
electron density of the carbon case rapidly decreases around $t=40$ fs.
}
\label{fig:fig-6}
\end{figure}

\begin{figure}[tbp]
\includegraphics[clip,width=6.5cm,bb=10 1 424 685]{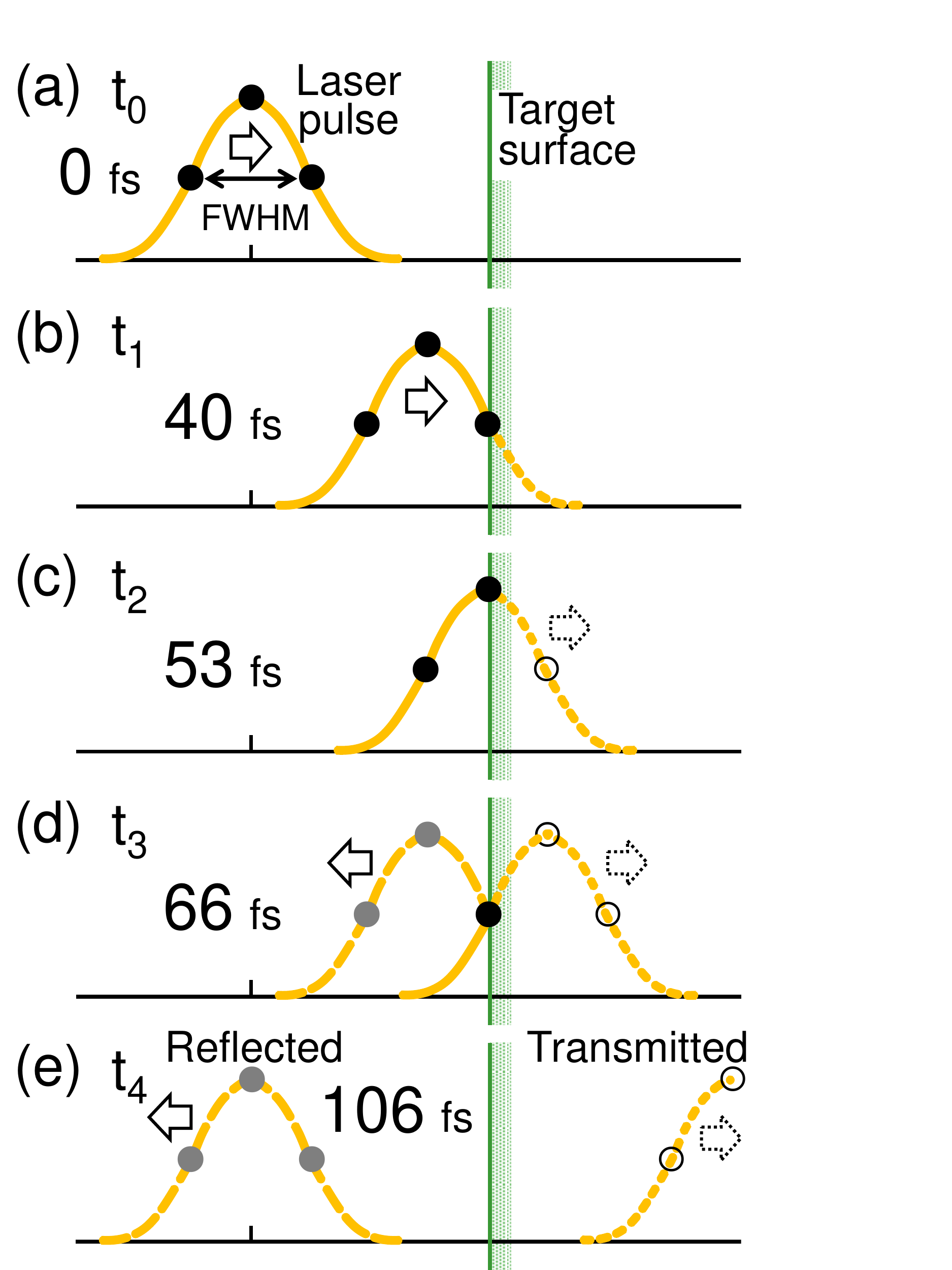}
\caption{
The relation between time and laser pulse position in our simulations
is shown by a simple diagram.
(a) Initial state of the laser pulse and the target, time $t_0=0$ fs.
(b) The FWHM point reaches the target surface at time $t_1=40$ fs.
(c) The center point of the laser pulse reaches the target surface
at time $t_2=53$ fs.
(d) The other FWHM point reaches the target surface at a time $t_3=66$ fs.
(e) The reflected laser returns to the initial position at a time $t_4=106$ fs.
}
\label{fig:fig-7}
\end{figure}

Figure \ref{fig:fig-6} shows target thickness and electron density
as functions of time for the $60 n_\mathrm{cr}$ carbon and gold cases.
The vertical axis of Fig. \ref{fig:fig-6}(a) is the normalized target thickness,
$\tilde{\ell}_x=\ell_x/\ell_0$, where $\ell_x$ is the thickness of the
first layer (see the inset in Fig. \ref{fig:fig-4}(b)),
and $\ell_0$ is the initial thickness.
Figure \ref{fig:fig-6}(b) presents the normalized electron density,
$\tilde{n}_e=n_e/n_0$, where $n_e$ is the electron density in the first layer
and $n_0$ is the its initial value.
The carbon target thickness becomes much greater than the gold target
for $t>40$ fs (Fig. \ref{fig:fig-6}(a))
because the half maximum point of the laser pulse on the $+x$ side is
located $12 \mu$m backward from the target (Fig. \ref{fig:fig-7}(a)), meaning
it arrives the initial target position at $t=40$ fs (Fig.\ref{fig:fig-7}(b)).
Therefore, the strong interaction between the target and the laser pulse begins
at this time, and the thickness of the carbon target starts to expand.
The center of the laser pulse arrives at the initial target position
at $t=53$ fs (Fig. \ref{fig:fig-7}(c)).
Around this time,
the interaction between the laser and the target is most strong.
The laser pulse reflected from the target surface arrives at the initial
position at $t=106$ fs (Fig. \ref{fig:fig-7}(e)).
Therefore, it is believed that the strong interaction between
the target and the laser pulse occurs during the interval $40<t<66$ fs.

The thickness of the carbon target increases beyond that of the gold target.
This difference becomes significant with time.
Thereby, the electron density in the carbon target starts rapidly
decreasing at around $t=40$ fs (Fig. \ref{fig:fig-6}(b)) and
then becomes much lower than that of the gold target for $t>40$ fs.
Even though half of the laser pulse remain at $t=53$ fs,
the electron density of the first layer differs greatly between the carbon
and gold cases. The electron density of the carbon target is about $50\%$
of the gold one, meaning that
the carbon target can more effectively absorb the remaining
laser pulse energy.

This is one of the reasons for which
we can obtain high energy protons in the carbon, ``light'' material, target;
``light'' material targets experience strong Coulomb explosions,
resulting in target thicknesses much greater than those of ``heavy'' materials
in the early stage of the interaction between the target and the laser pulse.
This means the electron density in the ``light'' material target becomes
much lower than that in the ``heavy'' material target when many parts of
the laser pulse remain without interacting with the target.
Therefore, the reflection of the laser pulse becomes small, and the absorbed
laser energy in the target increases in the ``light'' material target.
Note that this process occurs over a very short time;
the laser pulse duration is $27$ fs (FWHM) in our simulations.
In the ``heavy'' material target,
the reflected laser energy keeps increasing with density
since the target thickness does not much change during the interaction.
This means that the absorbed energy in the target continuously decreases.
Therefore, the obtained proton energy decreases with increasing target density.
Below, I present other reasons that high energy protons are obtained in
``light'' material targets and explain the occurrence of proton energy jumps.

\begin{figure}[tbp]
\includegraphics[clip,width=7.5cm,bb=10 7 490 420]{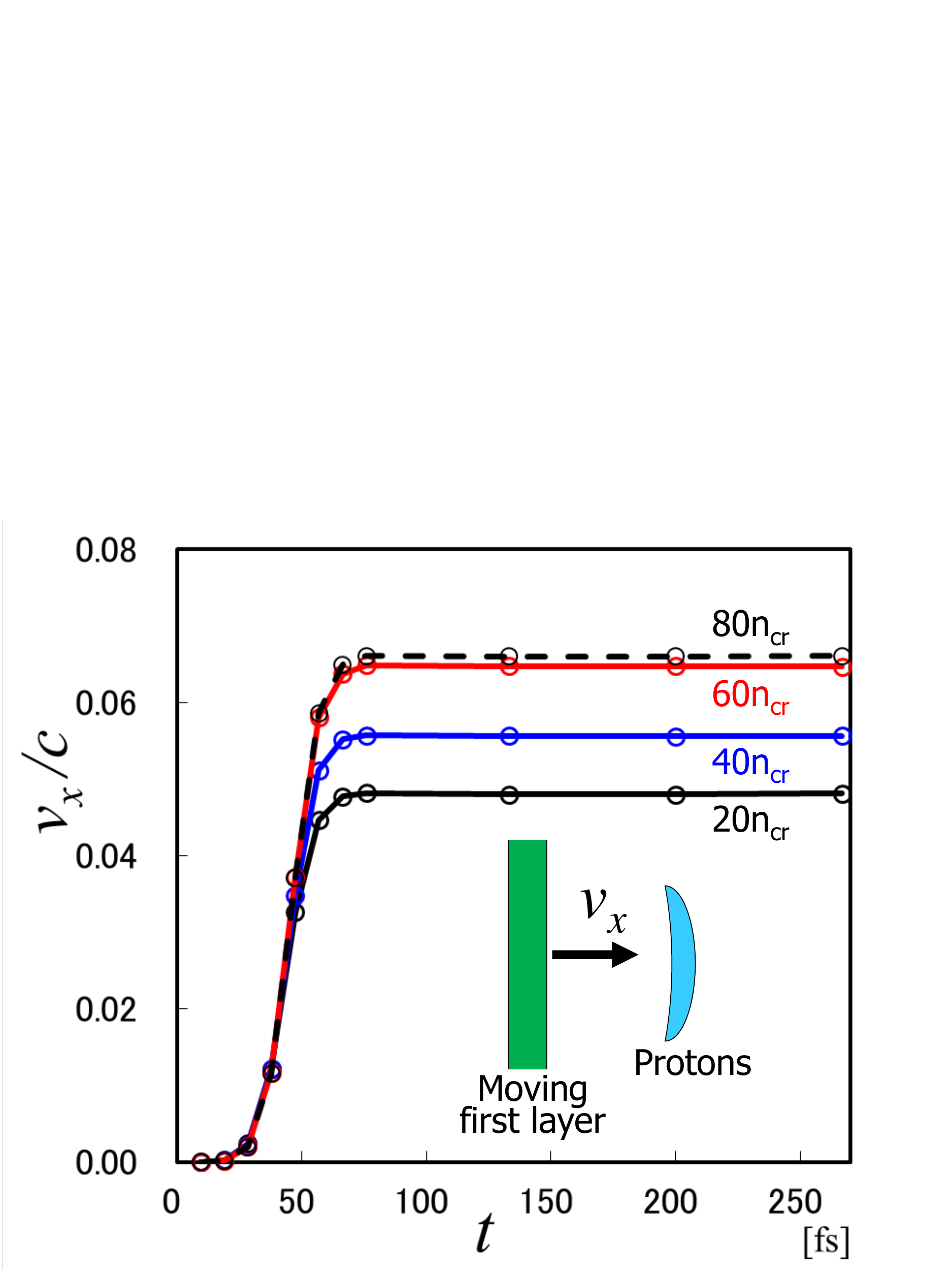}
\caption{
The first layer velocity in the $x$ direction normalized by the speed of light,
$v_x/c$, for each density, 20, 40, 60, and 80$ n_\mathrm{cr}$,
as a function of time for 3D simulations of the carbon case.
The first layer velocity, which is produced by RPA,
is faster in higher density targets.
}
\label{fig:fig-8}
\end{figure}

Figure \ref{fig:fig-8} shows the first layer velocity for the $x$ direction,
$v_x$, normalized by the speed of light, $c$, as a function of time,
for each carbon target density case in the 3D simulations.
The velocity is the averaged value for all ions in the first layer;
it rises rapidly in the initial period, $t=30 \sim 60$ fs,
when the laser pulse interacts with the target
(see Figs. \ref{fig:fig-4} and \ref{fig:fig-7}) and
then becomes constant for $t>70$ fs
after the laser pulse has been transmitted or reflects by the target.
The first layer velocities of the higher density target are faster than
those of the lower density target; this means strong RPA occurs in
the higher density cases because RPA occurs efficiently in targets that
are narrow and have high ion and electron densities, since those targets
form sharp separations of the high density ion and electron layers.
The moving first layer means that the moving electric potential for the
accelerating protons, efficiently accelerating them.
Therefore, the faster first layer generates higher energy protons,
i.e., high density targets can generate high energy protons.

\begin{figure}[tbp]
\includegraphics[clip,width=7.0cm,bb=7 0 416 364]{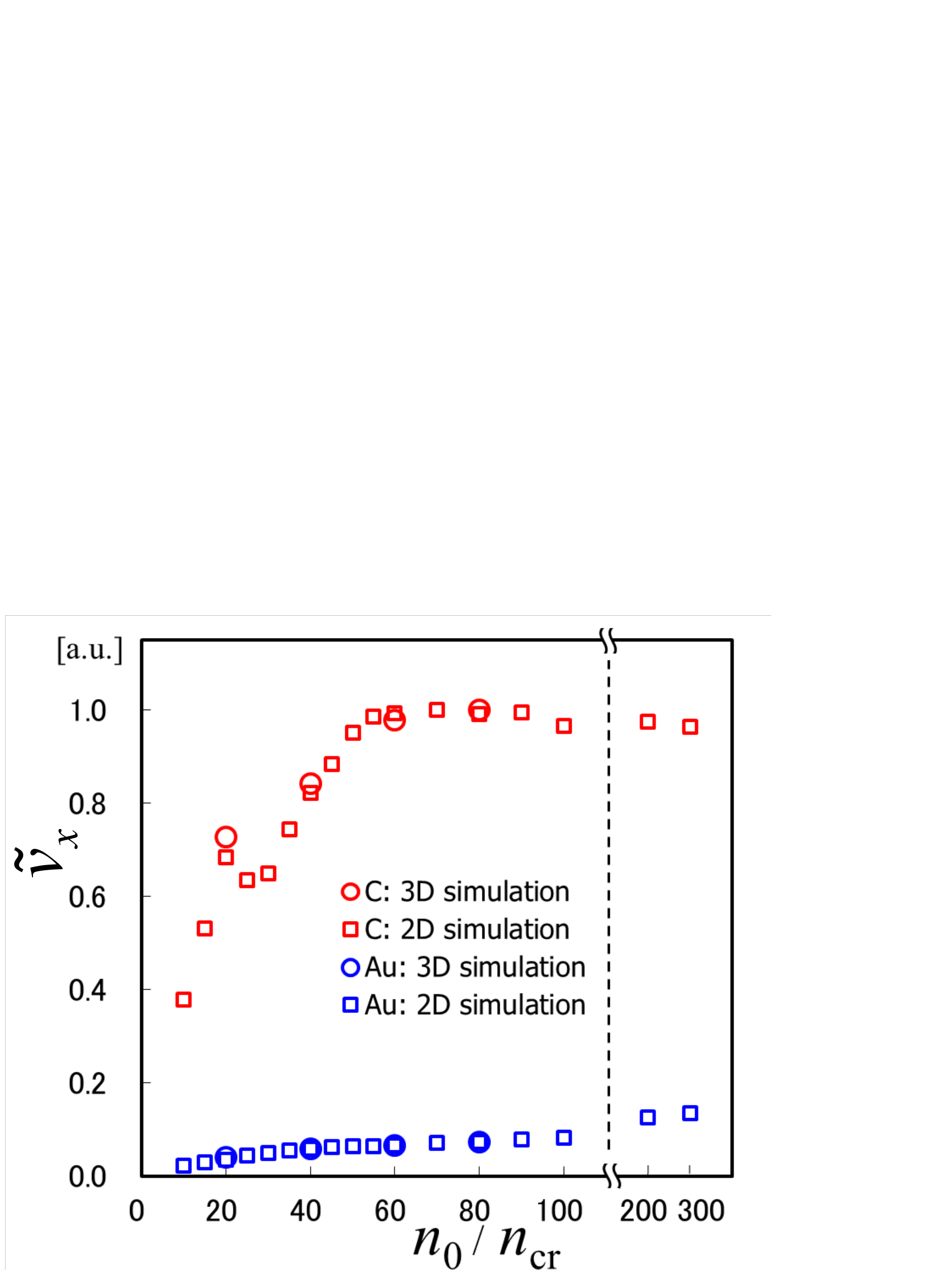}
\caption{
The first layer velocities versus first layer density at time $t=267$ fs.
The first layer velocities are normalized by the maximum value
in the carbon case.
The carbon target velocities are much higher than those in the gold case.
}
\label{fig:fig-9}
\end{figure}

Figure \ref{fig:fig-9} shows
the first layer velocities of the carbon and the gold cases as functions
of the first layer density at $t=267$ fs.
The velocities are normalized by the maximum value in both the 2D and 3D
carbon cases.
The carbon target velocities are much higher than those of gold at all
densities, and they rapidly increase until around $60 n_\mathrm{cr}$,
before becoming almost constant at $n_0>60 n_\mathrm{cr}$.
This trend is also the same as the figure of the obtained proton energy
(Fig. \ref{fig:fig-3}).
One of the reasons for the jump in the obtained proton energy
is RPA behavior.
On the other hand, the gold target velocity has a very small rise,
and the velocities are much smaller than those in the carbon case.
This means that the PRA effect is very small in the gold target.
This effect is significantly different between the carbon and gold cases.
The ``mass'' of the ions of targets plays an important role in RPA.
In the initial stage of the RPA process,
the electrons have almost the same momentum and distribution between the carbon
and gold cases, since they depend strongly on the laser conditions
and the initial electron distribution.
Therefore, the electric field working on the ions is also almost the same
in both cases.
The ``light'' ions then acquire high velocities, since they are ``light''.

The RPA behavior is another reason for which
high energy protons can be obtained from ``light'' material targets,
and help to explain the jump in the proton energy in such targets.
This process, RPA, starts at a very early stage of the ion acceleration
process and its duration is very short
(i.e., on the order of the laser pulse duration (FWHM),
which is 27 fs in our simulations).
Below, I present the other reason that high energy protons are obtained in
``light'' material targets and explain the occurrence of proton energy jumps.

\begin{figure}[tbp]
\includegraphics[clip,width=9.5cm,bb=35 86 490 482]{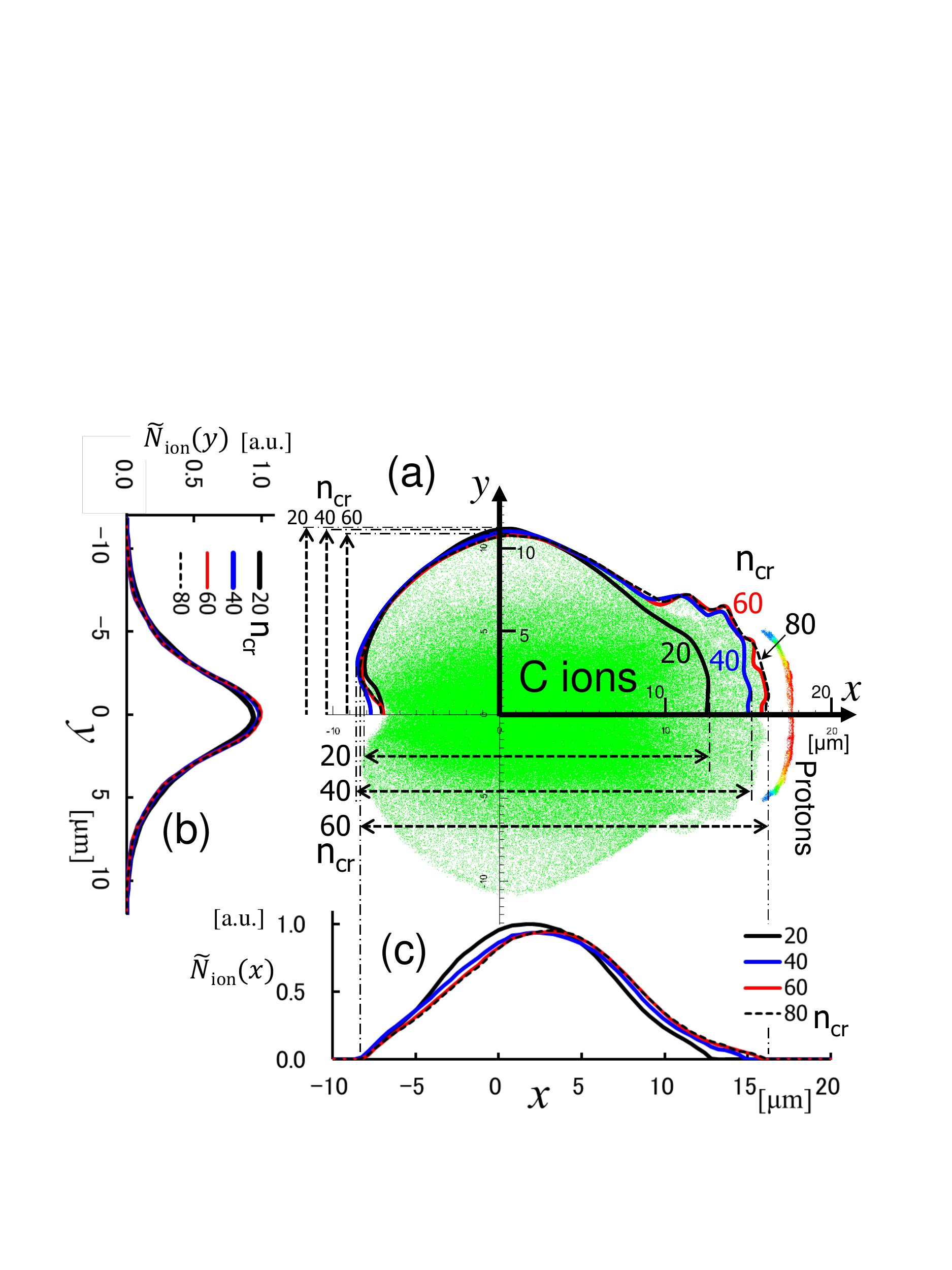}
\caption{
Spatial distribution of carbon ions at $t=200$ fs in the 3D simulation.
(a) Distribution of carbon ions (gas pattern in the center) and protons
(color scale corresponding to energy) in the $60 n_\mathrm{cr}$ case.
The solid and dotted lines show the outlines of the carbon ion cloud
at each density. 
The dotted arrows show the distribution area of ions in the $x$ and
$y$ directions.
(b) Distributions of the number of carbon ions in the $y$ direction,
and (c) in the $x$ direction.
$\tilde{N}_\mathrm{ion}$ is the number of ions per unit length along $x$, $y$,
and is normalized by its maximum value.
The ion cloud is distributed over a wider area in the $x$ direction with higher
density, although distributions in the $y$ direction are almost same in
all density cases. 
}
\label{fig:fig-10}
\end{figure}

Figure \ref{fig:fig-10} shows
a distribution of the ions in the $60 n_\mathrm{cr}$ carbon case at $t=200$ fs.
The carbon ion distribution by gas pattern and the protons by
color scale (Fig. \ref{fig:fig-10}(a)).
The ion distribution in the $-1\mu m<z<0$ region is shown as a projection
onto the $xy-$plane. Moreover, the outlines of the ion cloud
at each density case are shown by lines.
The ion cloud length in the $x$ direction, target thickness, is much longer
than the length in the $y$ direction, target radius.
This means that the Coulomb explosion in the target thickness direction
is harder than that in the radial direction.
Moreover,
the ions of the higher density targets are distributed over a longer area for
the $x$ direction,
and there is not much difference between the high density cases,
e.g., $60 n_\mathrm{cr}$ and $80 n_\mathrm{cr}$ (Fig. \ref{fig:fig-10}(a,c)).
On the other hand, in the $y$ direction, target radius direction, the width
of the ion distribution area is almost the same under all density cases
(Fig. \ref{fig:fig-10}(b)).
The carbon ions on the $+x$ side edge are close to the protons at this time,
meaning that the carbon ions keep pushing on the protons
and that this pushing force gets stronger with density as shown
in the high density cases, since the distance between them becomes shorter.
Although the maximum positions of ions in the carbon cloud for the $+x$
direction are more distant with higher density targets
(Fig. \ref{fig:fig-10}(a,c)), the positions on
the other side, $-x$ direction, are almost the same at all densities.
This is because
the velocity of the carbon ion cloud is higher under high ion density.
It is also shown that the PRA effect occurs strongly in thin and
high density targets.

Figure \ref{fig:fig-11} shows the ion cloud lengths in the $x$ and $y$
direction, $\ell_x$ and $\ell_y$ respectively, as functions of target density,
$n_0$, in the cases of carbon (Fig. \ref{fig:fig-11}(a)) and
gold targets (Fig. \ref{fig:fig-11}(c)).
For the carbon target, $\ell_x$ increases with target density,
becoming nearly flat above $60 n_\mathrm{cr}$
(similar to Fig. \ref{fig:fig-3}).
By contrast,
$\ell_y$ decrease almost monotonically with increasing target density,
meaning that the shape of carbon ion cloud for the high density target
is longer and narrower, an elongated ellipsoid.
On the other hand, in the gold target,
$\ell_x$ and $\ell_y$ are much shorter than in the carbon target,
and $\ell_x < \ell_y$, and are almost flat for the target density.
The same adjustment to the 2D results used Fig. \ref{fig:fig-5} is performed;
i.e., the 2D values are adjusted such that the averaged values at
$20, 40, 60,$ and $80 n_\mathrm{cr}$ would be the same as the 3D values.

Since the first layer of the target in our simulations is a narrow disk,
the target radius is much longer than its thickness.
The ions not around the edges of the target hardly move in the target radius,
$r$, direction, since there are many charged ions in the $+r$
and $-r$ directions.
In other words, the Coulomb forces in the $+r$ and $-r$ directions
are nearly balanced by those ions.
However, since there are few ions for the $x$ direction,
the ions move there comparatively easily.
Therefore, the target strongly expands in the $x$ direction, and
its tendency becomes stronger for high density targets than low density ones.
That is, the increased Coulomb force due to the increased ion density appears
remarkably in the $x$ direction in disk targets.
This effect increases along with density, ending
only above a certain density, $n_0>60 n_\mathrm{cr}$ in our conditions.
The stronger Coulomb explosion for the $x$ direction generates
higher energy protons.
This is the other reason for the jump in the proton energy and helps to
explain the high energy protons in high density targets of ``light'' material.

The contributions of the RPA and Coulomb explosion effects grow
for higher first layer densities.
The jump in the proton energy of the ``light'' material target is due to
the effect of the Coulomb explosion of the first layer and the effective RPA.
These are the reasons for obtaining high energy protons
in the ``light'' material targets.
The laser reflection from the ``light'' material target is smaller than that
of the ``heavy'' one, in spite of having the same electron and ion densities.
This also helps to explain why the obtained proton energies in the ``light''
material target are higher than those in the ``heavy'' one.
The ``heavy'' material targets have small
target expansion and RPA effects with a large amount of laser reflection.
Therefore, there is no energy jump as in ``light'' material targets and
the obtained proton energy is low.

\section{ANALYTICAL CONSIDERATION} \label{analy}

Here,
we analyze the RPA and Coulomb explosion effects.
Considerations are done using disk shaped targets.
This can be done with just disk targets such as the above simulations
or it is possible to consider
only the part where the laser irradiates the foil targets.

\subsection{RPA effect}

We consider the RPA effect for a relatively thick disk, i.e.,
nearly cylindrical.
It is assumed that the $x$-axis is in the cylinder height direction,
and the origin is at the middle of the cylinder whose
radius is $R$ and height is $\ell$.
The $x$ component of the electric field on the $x$ axis of the charged cylinder
is written as \cite{TM2}
\begin{equation}
E(x)=E_0 \left[a(x)+b(x) \right],
\label{exab}
\end{equation}
where
\begin{equation}
E_0=\frac{\rho\ell}{2\epsilon_0},
\label{e0}
\end{equation}
where $\rho$ is the charge density, $\epsilon_0$ is the vacuum permittivity,
and
\begin{equation}
a(x)=
\sqrt{\Bigl(\frac{x}{\ell}-\frac{1}{2}\Bigr)^2+\Bigl(\frac{R}{\ell}\Bigr)^2}
-\sqrt{\Bigl(\frac{x}{\ell}+\frac{1}{2}\Bigr)^2+\Bigl(\frac{R}{\ell}\Bigr)^2},
\label{exa}
\end{equation}
\begin{eqnarray}
b(x)=
\left \{ \begin{array}{ll}
-1 & (x < -\frac{\ell}{2}), \\
\frac{2x}{\ell} & ( -\frac{\ell}{2} \leq x \leq \frac{\ell}{2} ), \\
1 & (\frac{\ell}{2} < x).  \\
\end{array}
\right.
\label{exb}
\end{eqnarray}
Although $\rho$ and $\ell$ are variable in our simulations,
$\rho\ell$ is constant.
Therefore, $E_0$ is constant in all our simulation cases.
Here let us consider the situation where the electron layer moves
$\delta$ in the $+x$ direction maintaining its shape with respect
to the ion layer from the state in which those layers completely overlapped.
At this time, assuming that the charge density of the ion layer is $\rho$,
accordingly the charge density of the electron layer is $-\rho$.
The $x$ component of the electric field on the $x$-axis, $E(x)$,
in the ion layer $(-\ell/2 \le x \le \ell/2)$ is as follows.
When the ion and electron layers overlap over some area,
$0 \le \delta \le \ell$, the electric field, $E(x)$, is

for $-\ell/2 \le x \le -\ell/2+\delta$,
\begin{equation}
E_1(x)=E_0 \left[ a(x)-a(x-\delta) + \frac{2x}{\ell} + 1 \right];
\label{e1x}
\end{equation}

for $-\ell/2+\delta \le x \le \ell/2$,
\begin{equation}
E_2(x)=E_0 \left[ a(x)-a(x-\delta) + \frac{2\delta}{\ell} \right].
\label{e2x}
\end{equation}
When the ion and electron layers depart, $\ell < \delta$,

for $-\ell/2 \le x \le \ell/2$,
\begin{equation}
E_3(x) = E_1(x).
\label{e3x}
\end{equation}

Next, the force received by the ion layer is derived.
Within the ion layer,
let us consider the small region $\Delta V$ at position $x$ on the $x$-axis.
If the area of this region in the $yz-$plane is defined as $\Delta A$, and
the width in the $x$ direction is as defined $\Delta x$,
then $\Delta V = \Delta A \Delta x$.
The charge in this small region is
$q_{_{\Delta V}}=\rho\Delta V=\rho\Delta A\Delta x$, and
the force acting on this small region is
$F_{\Delta V} = q_{_{\Delta V}} E(x) = \rho\Delta A\Delta x E(x)$.
Let us now consider the force per unit area in the $yz-$plane,
$F_{\Delta x}= \rho E(x) \Delta x$.
The total force in the thickness direction of the ion layer is obtained
by integration in the $x$ direction:
\begin{equation}
F(\delta) = \int_{-\frac{\ell}{2}}^{\frac{\ell}{2}} F_{\Delta x} =
 \rho \int_{-\frac{\ell}{2}}^{\frac{\ell}{2}} E(x) dx.
\label{fdel}
\end{equation}
We substitute the formulas obtained above,
Eq. (\ref{e1x}), Eq. (\ref{e2x}), and Eq. (\ref{e3x}),
for $E(x)$ in this equation.

When the ion and electron layers overlap over some area,
$0 \le \delta \le \ell$,
\begin{eqnarray}
F(\delta)&=& \rho \Bigl[
  \int_{-\frac{\ell}{2}}^{-\frac{\ell}{2}+\delta} E_1(x) dx +
  \int_{-\frac{\ell}{2}+\delta}^{\frac{\ell}{2}} E_2(x) dx \Bigr] \nonumber \\
 &=& \rho \ell E_0 \Bigl[ \frac{1}{\ell} \int_{-\frac{\ell}{2}}^{\frac{\ell}{2}} [a(x)-a(x-\delta)] dx
 + \frac{2\delta}{\ell} - \frac{\delta^2}{\ell^2} \Bigr].
\label{fdel_a}
\end{eqnarray}

When the ion and electron layers depart, $\ell < \delta$,
\begin{eqnarray}
F(\delta)&=& \rho
  \int_{-\frac{\ell}{2}}^{\frac{\ell}{2}} E_3(x) dx \nonumber \\
 &=& \rho \ell E_0 \Bigl[ \frac{1}{\ell} \int_{-\frac{\ell}{2}}^{\frac{\ell}{2}} [a(x)-a(x-\delta)] dx
 + 1 \Bigr].
\label{fdel_b}
\end{eqnarray}
Therefore, if
$\int_{-\ell/2}^{\ell/2} [a(x)-a(x-\delta)] dx$
appearing in both expressions is evaluated,
$F(\delta)$ is given as a function of $\delta$.

In integrating this expression, we perform an approximate calculation
with accuracy preserved for the problem in which we are interested in.
Here, it is assumed the region in the $x$ direction to be studied
is relatively narrow, that is, it is in the vicinity of the cylinder
(i.e., $x$ is of the same order as or smaller than $\ell$),
and the height of the cylinder $\ell$ is smaller than its diameter, $2R$.
Thus, assuming $(\ell/2R)^2 \ll 1$ and $(x/R)^2 \ll 1$,
Eq. (\ref{exa}) is thus approximated as
\begin{equation}
a(x) \approx -\frac{x}{R}.
\label {aa1}
\end{equation}
In this paper, if $\epsilon \ll 1$, we take
$\epsilon^2,\epsilon^3,\cdots\approx 0$.

\begin{figure}[tbp]
\includegraphics[clip,width=9.0cm,bb=0 0 388 204]{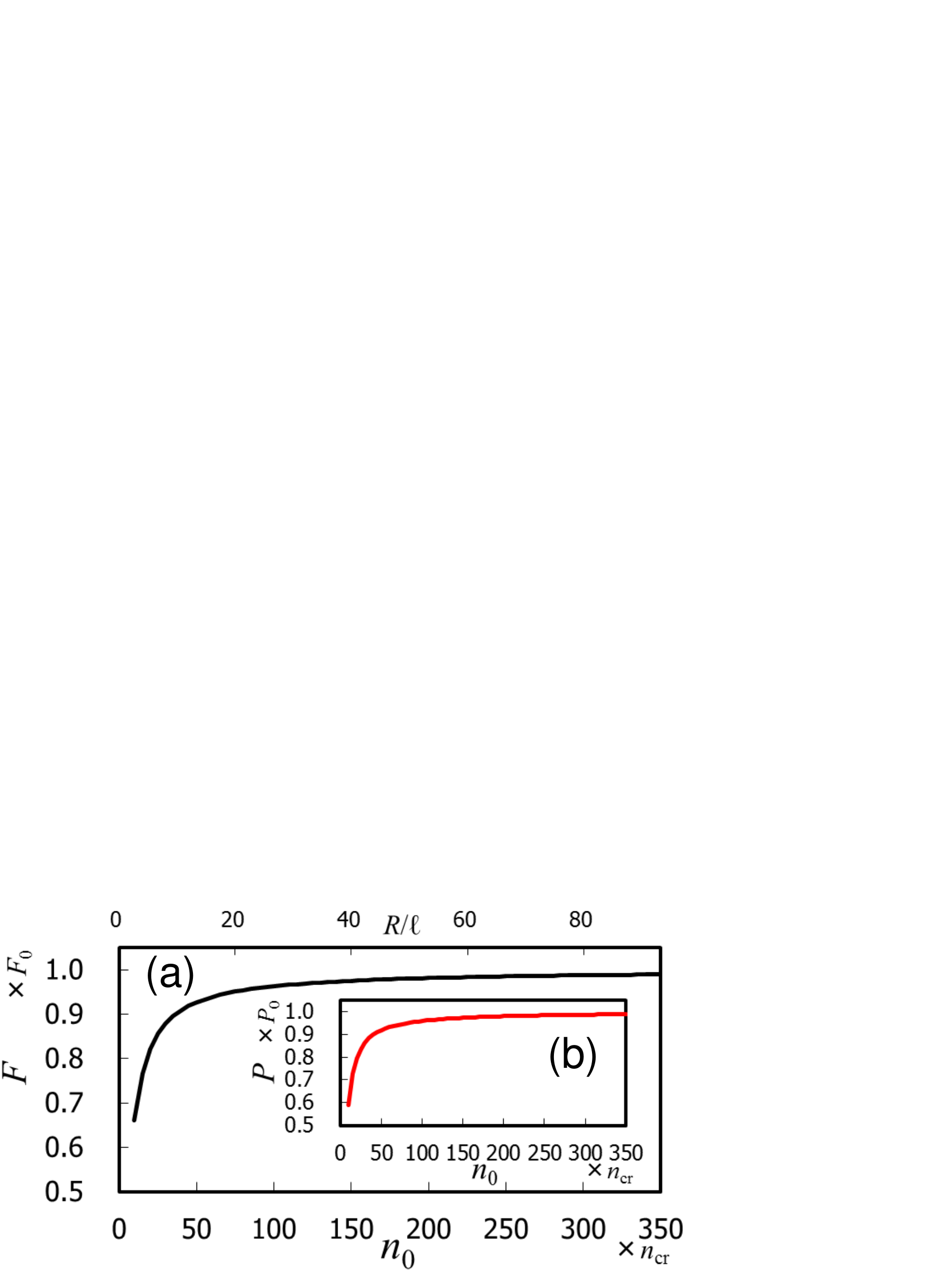}
\caption{
The maximum force, $F$, that the ion layer receives (a), and
the momentum, $P$, that ion layer obtains (b),
as functions of target density $n_0$
(the ion layer thickness, $R/\ell$, also shown).
Both jump until around $50n_\mathrm{cr}$.
}
\label{fig:fig-12}
\end{figure}

Therefore, $a(x)-a(x-\delta) \approx -\delta/R$.
Here, since $\delta$ must be on the same order as $x$,
it must satisfy $(\delta/R)^2 \ll 1$.
The integration can easily be performed, yielding
\begin{eqnarray}
F(\delta)=
\left \{ \begin{array}{ll}
F_0 [ -\frac{\delta^2}{\ell^2} +( \frac{2}{\ell} - \frac{1}{R})\delta ] & ( 0 \le \delta \le \ell), \\
F_0 ( 1-\frac{\delta}{R} ) & ( \ell < \delta ),
\end{array}
\right.
\label{fdel_c}
\end{eqnarray}
where
\begin{equation}
F_0 = \frac{(\rho \ell)^2}{2\epsilon_0}.
\label {f0}
\end{equation}
The force acting on the ion layer, $F(\delta)$, becomes maximal
when $\delta = \ell ( 1- \ell/2R )$; let this value be $F(\ell)$:
\begin{equation}
F(\ell)=F_0 \left( 1- \frac{\ell}{2R} \right) ^2.
\label {f0max}
\end{equation}
Thus, the thinner the target thickness, $\ell$,
the stronger the force that the ion layer receives.
Our condition is that $\rho\ell=$ constant;
therefore, if $\ell \to 0$, $\rho \to \infty$.
Fig.\ref{fig:fig-12}(a) shows the maximum force $F$ that the ion layer receives
for different target thicknesses $\ell$, and target densities $n_0$.
Here, $R=3.2\mu$m is used, which is the value of the previous simulations.
The force received by ion layer jumps to about $50n_\mathrm{cr}$
and becomes relatively flat afterward.
This shows the same trend as Fig. \ref{fig:fig-9}.

Above,
we discussed the maximum force, which is a static consideration.
Next, we consider the total force acting on the ion layer
during action time,
which is a dynamical consideration.
We evaluate the total force received by the ion layer while
the electron layer moves toward $x$ with respect to the ion layer from
the state in which the ion and the electron layers are completely overlapped.
The force acting on the ion layer
when the electron layer moves a distance $\delta$ with respect to the ion layer
is given by Eq. (\ref{fdel_c}).
Assuming that the electron layer moves at a constant speed $v$,
then $\delta=vt$ at time $t$.
Therefore, the force acting on the ion layer is
\begin{eqnarray}
F(t)=
\left \{ \begin{array}{ll}
F_0 [ -\frac{v^2t^2}{\ell^2} +( \frac{2}{\ell} - \frac{1}{R})vt ] & ( 0 \le t \le \frac{\ell}{v}), \\
F_0 ( 1-\frac{vt}{R} ) & ( \frac{\ell}{v} < t ).
\end{array}
\right.
\label{ft}
\end{eqnarray}
Here, it must satisfy $(vt/R)^2 \ll 1$.

Next, we derive the sum of $F(t)$ up to time $T$, 
that is, the ion layer's obtained momentum, $P(\ell)$.
The time, $T$, is taken when the electron layer is near the ion layer,
although the electron layer is departed ($T>\ell/v$).
It is because we see that the RPA appears at the initial acceleration stage
in our simulation results,
i.e., the electron layer expands widely as it goes away from the ion layer
and the RPA effect then remarkably decreases.
The momentum, $P(\ell)$, is
\begin{eqnarray}
P(\ell)&=&
  \int_{0}^{T} F(t) dt
 = \int_{0}^{\frac{\ell}{v}} F_1(t) dt + \int_{\frac{\ell}{v}}^{T} F_2(t) dt \nonumber \\
 &=& F_0 \Bigl( -\frac{\ell}{3v} + T - \frac{vT^2}{2R} \Bigr),
\label{fdel-b}
\end{eqnarray}
where $F_1(t)$ is the upper equation in Eq. (\ref{ft}), and
$F_2(t)$ is the lower equation.
It must satisfy $(\ell/2R)^2 \ll 1$, and $(vT/R)^2 \ll 1$.
In all our simulation cases,
$\rho\ell$ is constant, therefore $F_0$ is also constant.
Therefore, $P(\ell)$ takes the maximum value $P_0$ when $\ell=0$,
\begin{equation}
P_0 = F_0 T \left( 1-\frac{vT}{2R} \right).
\label{p-max}
\end{equation}
$P(\ell)$ is expressed using $P_0$,
\begin{eqnarray}
P(\ell) &=& P_0 - \frac{\ell}{3v}F_0  \nonumber \\
        &=& P_0 \left[ 1-\frac{\ell}{3vT(1-\frac{vT}{2R})} \right].
\label {ptil}
\end{eqnarray}
For smaller target thickness $\ell$, resulting in larger $\rho$,
the momentum that the ion layer gains is bigger.

From here on, $T$ is taken to be the time at which the ion and the electron
layers separate into exactly two layers, both in contact with each
other's surfaces, in the case of the longest $\ell$ of our simulations.
This corresponds to $10n_\mathrm{cr}$.
Let $\ell_{10}$ be the target thickness at $10n_\mathrm{cr}$;
then, $T=\ell_{10}/v$, and therefore
$P(\ell)=P_0[1-\ell/3\ell_{10}(1-\frac{\ell_{10}}{2R})]$.
Fig.\ref{fig:fig-12}(b) shows the change of $P$ with respect to the change of
$\ell$, indicating a change of $n_0$.
The momentum obtained by the ion layer jumps significantly around
$50n_\mathrm{cr}$. It has also the same trend as Fig. \ref{fig:fig-9}.

The above considerations explain why the obtained proton energy jumps
to about $50n_\mathrm{cr}$ in the RPA regime under our conditions.

\subsection{Target expansion effect}

In the simulation, the high density disk expands strongly
in the $x$ direction due to its Coulomb explosion,
although there is almost no difference in radial expansion.
Here, we consider the difference in disk expansion due to
the difference in density.

The $x$-axis is defined as in the previous subsection and
the $y$ and $z$-axes are parallel to the disk surface.
In our conditions,
the total numbers of ions, i.e., electrons, is the same in all cases.
The density is changed by changing the disk thickness,
and the disk has the same shape in the radial direction, in the $yz-$plane.
Now, let disk-1 have thickness $\ell$, area $A$, and ion density $\rho_1$.
Let disk-2 be denser than disk-1 and let disk thickness be $\alpha\ell$.
At this time, disk-2 is formed by compressing disk-1 in the $x$ direction;
therefore, $0 <\alpha <1$.
The positions of ions are assumed to change only along the $x$ coordinates,
and the $y$ and $z$ coordinates of disks-1 and -2 are the same.

Let $N$ be the total number of ions in the disk.
Since the volume of disk-1 is $V_1=A\ell$,
the charge density is $\rho_1=Nq/A\ell$,
where $q$ is the ion charge,
and the surface charge density is $\sigma_1=\rho_1\ell$.
On disk-2, $V_2=A\alpha\ell$; therefore, $\rho_2=Nq/A\alpha\ell=\rho_1/\alpha$,
and the surface charge density is
$\sigma_2=\rho_2\alpha\ell=\rho_1\ell=\sigma_1$.
That is, when the disk is projected onto the $yz-$plane,
the area density of the ions on disks-1 and -2 are the same,
and the ion distribution in the plane is also the same for both disks
according to the above definition. 

On the other hand,
considering the case where ions are projected onto the $x$-axis in the same way,
the linear charge density of disk-1 is $\lambda_1=Nq/\ell$, and that of
disk-2 is $\lambda_2=Nq/\alpha\ell=\lambda_1/\alpha$.
Therefore, the linear density differs between both disks
and is $\lambda_2>\lambda_1$ in the case where
the plate thickness of disk-1 $>$ disk-2.

Therefore, the distance between each of the ions
in the disk radial direction, $r$, is almost the same for both disks,
i.e., the radial Coulomb force acting between each of the ions is almost
the same for both disks.
On the other hand, in the thickness direction, $x$ direction, of the disk,
the distance between ions is different in both cases,
and that for disk-2, the high density case,
is shorter than disk-1, the low density case.
Therefore, the force in the $x$ direction acting between ions is
larger for disk-2.
Thus,
the Coulomb expansion in the thickness direction becomes larger
in the case of high density,
although explosion in the radial direction is almost the same.
This is discussed in detail below.

In the previous section,
since an electric field inside the disk and near its surface is required,
a suitable approximate expression is used.
Here,
since the expression for the electric field at a position where
near the disk surface and
away from the disk surface in the $x$ direction is required,
and it is not necessary the one where the inside of the disk,
we obtains a suitable approximation.
Here, it is assumed that disk is thin compared to its diameter,
i.e.,
using the condition of $\ell/2R \ll 1$, Eq. (\ref{exa}) is thus approximated as
\begin{equation}
a(x) \approx -\frac{x}{\sqrt{x^{2}+R^{2}}}.
\label{axx2}
\end{equation}
Therefore, the electric field on the $x$ axis is
\begin{equation}
E(x)= E_0\left(1-\frac{x}{\sqrt{x^{2}+R^{2}}} \right).
\label{exx}
\end{equation}
It must be $x \geq \ell/2$ in here.

The model used below is presented here.
Let the total number of ions contained in the disk be $N$, and no electrons.
It is assumed that the disk thickness, $\ell$, is equally divided into $N$
layers of thickness $\Delta \ell=\ell/N$,
with only one ion existing in each.
The $y$ and $z$ positions of the ion are arranged such that ions are evenly
distributed in the $yz-$plane when projected there in all cases.

According to the above rule,
the energy needed to generate the disk with $N$ ions
by assembling them one by one from $x=+\infty$ is shown here.
Consider the situation where $n$ ions have already been assembled.
At this stage,
the disk radius is $R$, and the thickness is $\ell_n=n\Delta\ell=n\ell/N$.
We find the energy to bring $1$ ion from $x=+\infty$ onto the surface of
this disk.
This means adding $1$ layer with a thickness of $\Delta\ell$ on the disk.
If the end point of the ion is decided,
the energy is the same regardless of route.
Therefore, we use the following method.
First,
we bring an ion from $+\infty$ along the $x$-axis to the next layer position.
Next, we move the ion in the radial, $r$, direction of the disk,
such that the ion distribution in the $yz-$plane becomes uniform.
This operation is performed on $1$ to $N$ ions, and
the energy required to bring these ions is calculated.
Since $E_0=\rho l_n/2\epsilon_0=qn/2\pi\epsilon_0 R^2$,
the electric field of a disk with $n$ ions is
\begin{equation}
E_n(x) = \frac{qn}{2\pi\epsilon_0 R^2}
       \left(1-\frac{x}{\sqrt{x^{2}+R^{2}}} \right).
\label{exx2}
\end{equation}
Therefore,
the energy required to bring $1$ ion along the $x$-axis from $+\infty$ to
the disk surface, $x=\ell_n/2$, is
\begin{eqnarray}
\mathcal{E}_x(n) &=& \int_{\frac{\ell_n}{2}}^{\infty} qE_n(x) dx \nonumber \\
 &\approx& \frac{q^2}{2\pi\epsilon_0 R^2} \left( \frac{\ell^2}{8RN^2}n^3 - \frac{\ell}{2N}n^2 + Rn \right).
\label{en1}
\end{eqnarray}
$(\ell_n/2R)^2 \ll 1$ is used for this purpose.
Let $\mathcal{E}_r(n)$ be the energy required to move an ion,
which has been brought to the center of the $\Delta\ell$ layer,
in the radial direction, $r$, within the $\Delta\ell$ layer.
Then, the energy required to place an ion at a setting position
on the disk surface is $\mathcal{E}(n)=\mathcal{E}_x(n)+\mathcal{E}_r(n)$.
The energy required to bring $1$ to $N$ ions is
$U(\ell)= \int_1^N \mathcal{E}(n)dn=U_x(\ell)+U_r(\ell)$,
where
\begin{eqnarray}
U_x(\ell) &=& \int_1^N \mathcal{E}_x(n) dn \nonumber \\
  &\approx& \frac{q^2N^2}{2\pi\epsilon_0 R^2} \left( \frac{\ell^2}{32R} -
  \frac{\ell}{6} + \frac{R}{2} \right).
\label{uuu}
\end{eqnarray}
It is used the condition of $N \gg 1$.
$U_x(\ell)$ has a minimum value of $\ell=8R/3$ and
a decreasing function in $\ell<8R/3$.
Therefore, $U_x(\ell)$ has a maximum value at $\ell=0$ and
\begin{equation}
\mathrm{max} \{ U_x(\ell) | 0 \leq \ell \leq 8R/3 \} = U_x(0) = \frac{q^2N^2}{4\pi\epsilon_0 R}.
\label{umax}
\end{equation}
We call this value $U_0$.
Since $qN=\rho\ell\pi R^2$,
\begin{equation}
U_0 = \frac{(\rho\ell)^2\pi R^3}{4\epsilon_0}.
\label{u0}
\end{equation}
Equation (\ref{uuu}) is written as
\begin{equation}
U_x(\ell) = U_0 \left( \frac{\ell^2}{16R^2}-\frac{\ell}{3R} + 1 \right).
\label{uu2}
\end{equation}
And $U_r(\ell)= \int_1^N \mathcal{E}_r(n)dn$.

\begin{figure}[tbp]
\includegraphics[clip,width=8.0cm,bb=7 0 390 205]{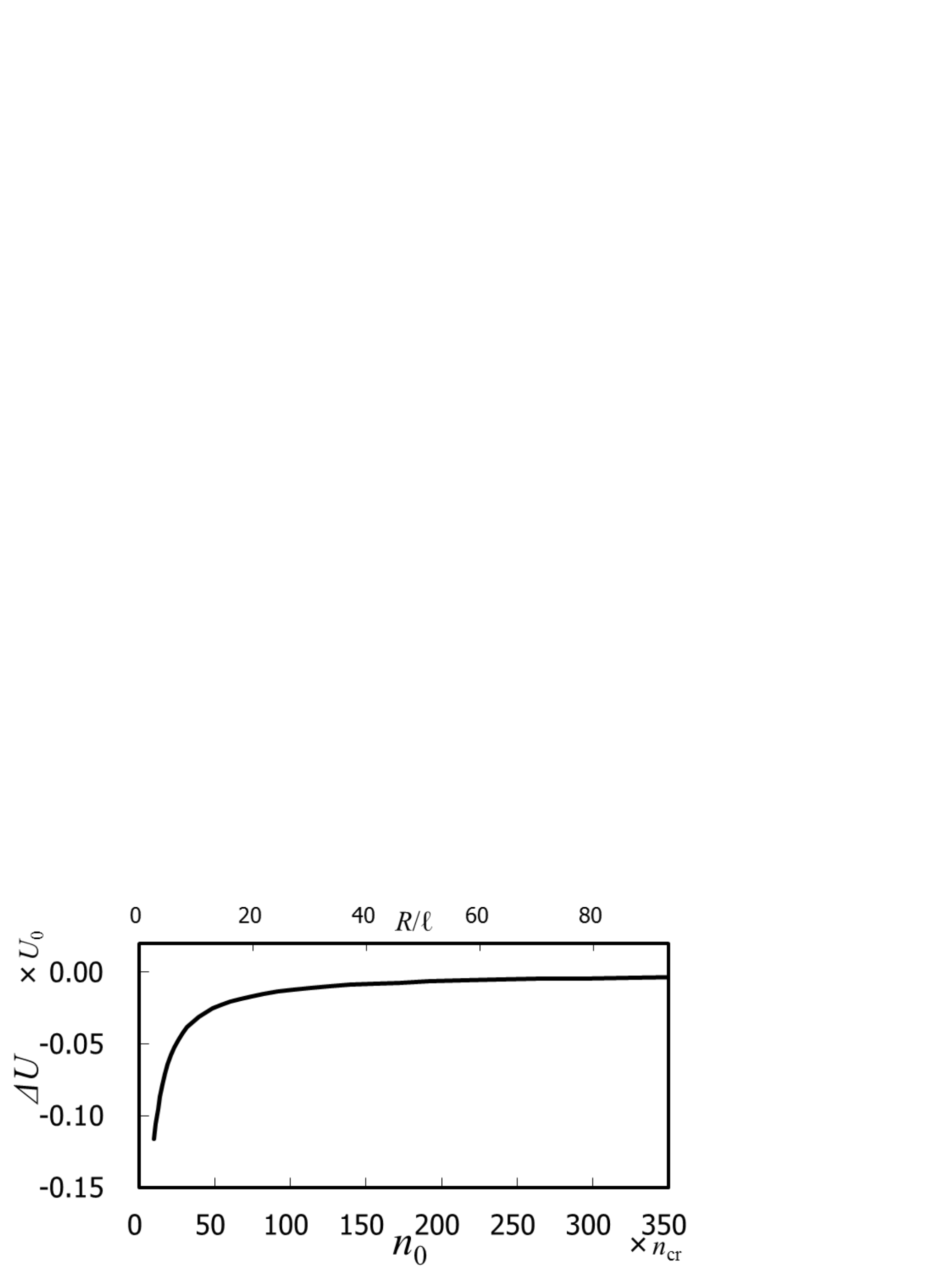}
\caption{
The differential energy of the ion layer, $\Delta U$,
as a function of target density $n_0$
(the ion layer thickness, $R/\ell$, also shown).
It jumps until around $50n_\mathrm{cr}$.
}
\label{fig:fig-13}
\end{figure}

Since we want to know the change in energy, $\Delta U$,
due to the change of the disk thickness, $\ell$,
we take the difference from the base energy, $U(0)$,
that is energy at $\ell=0$, which is the maximum value.
That is, $\Delta U=U(\ell)-U(0)=U_x(\ell)-U_x(0)+U_r(\ell)-U_r(0)$.
We assume that the ion positions in the $yz-$plane are the same for
different thicknesses and that the disk is thin, $\ell \ll 2R$;
therefore, $U_r(\ell)\approx U_r(0)$.
We obtain
\begin{equation}
\Delta U(\ell) = U_0 \left( \frac{\ell^2}{16R^2}-\frac{\ell}{3R} \right).
\label{duu}
\end{equation}
Figure \ref{fig:fig-13} shows the change of $\Delta U$ with respect to
those of $\ell$, and $n_0$.
The energy, $\Delta U$, jumps significantly until around $50n_\mathrm{cr}$.

\section{CONCLUSIONS}
Proton acceleration driven by a laser pulse irradiating a thin disk target is
investigated with the help of 2D and 3D PIC simulations.
The obtained proton energy change is investigated as a function of the density
changes of ``light'' and ``heavy'' material targets.
The obtained proton energy jumps at a certain density of the first layer
using a ``light'' material.
This is caused by the Coulomb explosion, RPA,
and the reflection of the laser pulse.
On the other hand,
such jumps did not occur in the ``heavy'' material target.
Higher energy protons could thus be obtained using a ``light'' material target
above a certain density;
this is because, as seen in our simulations, these targets experience
strong Coulomb expansion, efficient RPA, and less laser reflection.
Materials that can easily become ``light'' are ideal for the target
since they can be almost fully striped at low energy,
during the initial stage of laser pulse irradiation.
This is because
the target expansion of such materials by a Coulomb explosion
can start at a very early stage in the acceleration process,
and RPA acts in the initial stage in the acceleration process.
Therefore,
it is effective for ``lighter'' ions to form quickly,
before the main part of the laser pulse arrives.
This means an ordinary light material.
That is, light materials are more effective than heavy ones
for the target to obtain high energy protons.

In general, there is a pre-pulse before the main pulse.
As seen in our simulations, if the density of the expanded target
from the pre-pulse is over a certain value,
the energy of the obtained ions is almost the same.
The density is about $60 n_\mathrm{cr}$ under our simulation conditions,
which is ordinary carbon that has expanded by about 6 times.
That is, if we reduce the pre-pulse to that level,
it has little effect on the obtained ion energy.

\section*{ACKNOWLEDGMENTS}
I thank S. V. Bulanov, T. Zh. Esirkepov, M. Kando, J. Koga, and K. Kondo
for their useful discussions.
The computations were performed using the ICE X supercomputer at JAEA Tokai.

\end{document}